\documentclass{emulateapj}
\newcommand{\msuns}{M$_{\sun}~$}

\begin{document}

\title{EXTRASOLAR PLANET ECCENTRICITIES FROM SCATTERING \\ IN THE PRESENCE OF RESIDUAL GAS DISKS}

\author{Nickolas Moeckel\altaffilmark{1}, Sean N. Raymond\altaffilmark{1,2}, Philip J. Armitage\altaffilmark{3}}
\affil{ Department of Astrophysical and Planetary Sciences \\
	University of Colorado, Boulder, CO 80309}
	\altaffiltext{1}{CASA, University of Colorado, Boulder, CO 80309}
	\altaffiltext{2}{NASA Postdoctoral Program Fellow}	
	\altaffiltext{3}{JILA, University of Colorado, Boulder, CO 80309}
\email{moeckel@colorado.edu}

\begin{abstract}
Gravitational scattering between massive planets has been invoked to explain
the eccentricity distribution of extrasolar planets. For scattering to occur,
the planets must either form in -- or migrate into -- an unstable
configuration. In either case, it is likely that a residual gas disk, with a
mass comparable to that of the planets, will be present when scattering
occurs. Using explicit hydrodynamic simulations, we study the impact of gas
disks on the outcome of two-planet scattering. We assume a specific model in
which the planets are driven toward instability by gravitational torques from
an outer low mass disk. We find that the accretion of mass and angular
momentum that occurs when a scattered planet impacts the disk can strongly
influence the subsequent dynamics by reducing the number of close
encounters. The eccentricity of the innermost surviving planet at the epoch when 
the system becomes Hill stable is not
substantially altered from the gas-free case, but the outer planet is
circularized by its interaction with the disk. The signature of scattering
initiated by gas disk migration is thus a high fraction of low eccentricity
planets at larger radii accompanying known eccentric planets. Subsequent 
secular evolution of the two planets in the presence of damping can further 
damp both eccentricities, and tends to push systems away from apsidal alignment 
and toward anti-alignment. We note that the late 
burst of accretion when the outer planet impacts the disk is in principle 
observable, probably via detection of a strong near-IR excess in systems 
with otherwise weak disk and stellar accretion signatures.  
\end{abstract}

\keywords{accretion, accretion disks --- scattering --- planets and satellites: formation ---
planetary systems: formation --- planetary systems: protoplanetary disks}

\section{MOTIVATION}
The eccentricity distribution of massive extrasolar planets beyond the tidal
circularization radius is broad, with a median eccentricity $<e> \simeq 0.25$
\citep{marcy05}. Since giant planet formation models predict that the envelope
is accreted when the planet has an almost circular orbit, the observed
eccentricities are generally interpreted as evidence of evolutionary processes
that occur subsequently. One possibility is that the same gravitational
torques from the gas disk that drive orbital migration also excite
eccentricity
\citep{artymowicz91,ogilvie03,goldreich03,moorhead08}. Alternatively, the
eccentricity may arise from gravitational instability within an unstable
multiple system \citep{rasio96,weidenschilling96,lin97}. {\it N}-body
experiments show that a wide variety of unstable initial conditions -- which
can include systems with two, three or many giant planets -- evolve under
scattering to yield an eccentricity distribution similar to that observed
\citep{chatterjee07,juric07}. The agreement between theory and observations
supports the scattering scenario, though other possibilities are not ruled
out.

To date, most studies of planet-planet scattering have considered purely
gravitational evolution from gas-free initial conditions. Gas has been ignored
primarily for reasons of computational necessity, since fairly robust
arguments suggest that at least {\em small} amounts of gas are likely to
survive to the epoch when scattering occurs (planetesimal disks will also be
present, though these will likely be dynamically important only in systems
similar to the Solar System, in which at least some giant planets at large
radii have low envelope masses). In particular, for two-planet systems most
initial separations within the boundary that is Hill stable for all time
\citep{gladman93} are unstable on very short time scales.  Realistically, an
unstable two planet system is therefore likely to arise only if the planets
are driven toward such a state, for example due to gas disk
migration. Efficient migration requires a disk mass that is of the same order
as the planet mass. Multiple planet systems with $N_p > 2$ exhibit broader
instability regions, but even in this case the most likely scenario is that
instability sets in once eccentricity damping due to the presence of gas
becomes small enough. Again, this will occur once (to order of magnitude) the
gas disk mass falls to the level of the planet mass, and scattering is likely
to occur in an environment that is gas poor but not gas free.

In this paper we use hydrodynamic simulations to study the influence of low
mass residual gas disks on the outcome of gravitational scattering. As noted
above, there are many possible scenarios that yield different initial
conditions for scattering experiments -- here we adopt a simple model that is
computationally tractable. We assume that two Jupiter-mass planets form on
well-separated circular orbits. Models of giant planet formation suggest that 
the onset of runaway accretion occurs when the cores attain a mass of 10~$M_\oplus$ 
\citep{hubickyj05}, and, since neighboring cores are likely 
to form with a separation of $\sim 10$ mutual Hill radii \citep{kokubo98}, this suggests that 
plausible initial separation ratios are $\simeq 1.3$ or greater. If  
the gas disk interior to the outer planet is relatively weak migration 
driven by the residual outer gas results in convergence of the orbits. The 
initial separation alluded to above lies close to the 3:2 resonance, so depending 
upon the precise initial separation convergent migration might result in resonance 
crossing. If the resonance {\em is} encountered, the result can either be trapping 
into resonance \citep{lee02,snellgrove01,kley05}\footnote{The existence of a number 
of known resonant pairs of planets supports the idea that trapping occurs in at 
least some cases.} or, if the disk is turbulent \citep{adams08}, 
avoidance of trappping and ongoing convergence of the orbits. In the latter case 
the outer planet continues to 
migrate inward on a viscous time scale until the 
separation of the planets is driven past the Hill stability criterion, and a scattering
event occurs. The goal of this paper is to answer the question: what 
effect does the remnant disk now have on the scattering
planetary system?

\section{SIMULATION METHOD AND INITIAL CONDITIONS}
Hydrodynamic simulations of the disk over dynamically interesting time scales
are computationally difficult. In this preliminary study we consider only a
small number of cases, and set up the initial conditions so that scattering
occurs early in the simulation. Using purely {\it N}-body runs, we first
identified a large number of initial conditions in which two Jupiter mass
planets, orbiting a Solar mass star on nearly circular orbits separated by 2.5
to 3 mutual Hill radii, are about to experience a first close encounter. For
realistic migration rates \citep{armitage07} instability is likely to set in
at approximately this separation. The orbital radius of the inner planet was
set to 3 AU.  The planets were nearly coplanar, with small inclinations $i \lesssim 1^\circ$.  We chose five initial conditions from these {\it N}-body runs spanning a range in the
closest approach distance attained during the first encounter, summarized in Table~\ref{sim_params}.  

\begin{deluxetable}{cccccc} 
\tablecolumns{6} 
\tablewidth{0pc} 
\tablecaption{Simulation parameters} 
\tablehead{ 
\colhead{Case} & \colhead{$d_{close}$} 
&\colhead{$m_{out,f}$}
& \colhead{$L_{acc}^{max}$}  &\colhead{$<L_{acc}>$}  &\colhead{$\zeta_{f}$} \\
\colhead{} & \colhead{(AU)} 
& \colhead{$(M_J)$} 
& \colhead{$(10^{32}erg~s^{-1})$}  &\colhead{$(10^{32}erg~s^{-1})$} &\colhead{$(\zeta_{crit})$}
}
\startdata 
A & 0.00094 & 1.48 & 6.26 & 1.76 &1.027 \\ 
B & 0.0075  & 1.54 & 11.37 & 2.54 &1.040 \\ 
C & 0.044 & 1.40 & 11.51 & 2.75&1.005 \\ 
D & 0.11& 1.52 & 18.82 & 4.68&1.034 \\ 
E & 0.30  & 1.61 & 18.36 & 5.82&1.045 \\
%x 1.11 1.08 1.08 1.06  <- m_in values
\enddata
\tablecomments{The closest approach during the initial scattering is $d_{close}$.  Subscript $f$ refers to the quantity at the end of the simulations, $10^4$ yr.  $<L_{acc}>$ is the mean accretion luminosity for the first 500 years of the simulation.}
\label{sim_params}
\end{deluxetable}

We evolve the initial conditions for $10^4 \ {\rm yr}$ both as a purely {\it N}-body system, and hydrodynamically
including the influence of an external gas disk.  The remnant disk is assumed to extend from 5-10~AU with a
surface density profile $\Sigma(r) = \Sigma_0 r^{-3/2}$, normalized so that the gas disk contains one Jupiter
mass of gas.  This setup is quite similar to the scenario considered by \citet{moorhead05}, in which two-planet
systems that had cleared an inner gap were driven toward instability via Type II migration of the outer planet. 
These authors include the effect of a remnant disk on planet-planet scattering via an eccentricity damping
prescription.  Likewise, \citet{chatterjee07} studied the damping effect of a disk on three-planet scattering by
coupling a planet-disk torque prescription with a 1-D disk model. In most of our models 
we assume that the inner disk is weak enough as to be neglected. In the regime we consider 
considerable depletion of the inner disk is likely, since the angular momentum of the planets 
is large enough to significantly retard the inflow of the outer disk while the inner disk continues 
to accrete on the normal viscous time scale. However, full depletion of the inner disk is only 
an approximation\footnote{Unless magnetic effects \citep{chiang07} or photoevaporation 
\citep{clarke01} operate and clear a true cavity out as far as the radius of the inner planet.}, 
and recent calculations suggest that some gas will either survive in the inner region or 
flow past the planets from the outer disk to resupply the inner disk \citep{crida07,lubow06}. 
We therefore ran one additional simulation to ascertain the effect of a small mass of gas 
interior to the planets on the results. Using the results of \citet{lubow06}, we estimated that 
reasonable amounts of material in the depleted inner disk will likely be of the order 10\% of the initial disk
profile \citep{lubow06}, so that the inner disk has $\Sigma(r) = 0.1 \Sigma_0 r^{-3/2}$.  If the inner disk
extends from 0.75 - 2 AU, it contains $\sim 0.06 M_J$ of material, much less than the outer disk.  If the inner
planet scatters into this inner remnant a small amount of eccentricity damping and mass loading onto the planet
could occur.

\begin{figure}
\centering
  \plotone{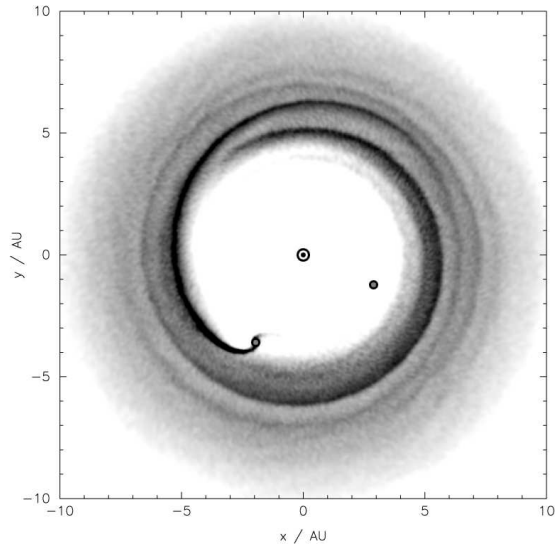}
  \caption{An example snapshot from case C illustrating the interaction between the outer planet and the disk.  Plotted \citep[using the program {\tt splash,}][]{price07} is the column density, with the star ({\it dotted circle}) at the center and the two planets ({\it filled circles}) in orbit.  The outer planet has induced density waves in the disk, from which it is accreting gas.  This snapshot is $\sim$ 30 years after the close encounter between the planets.}
  \label{density_example}
\end{figure}

We evolve the system using the GADGET-2
smoothed particle hydrodynamics (SPH) code \citep{springel05}, modified so
that the star and planets behave as sink particles \citep{bate95}.  The
accretion radius of the planets is set to $\sim 50R_J$.  The star and planets
interact with each other and the gas particles only via gravity, which is
effectively unsoftened.  A further modification to the code was made to ensure
that the force between the star and planets is computed directly, and not with
a tree-code approximation. For the runs presented here the disk is represented
by $\sim 1.75 \times 10^5$ particles.  At this resolution the Hill radius of
the planets is resolved by several SPH smoothing lengths throughout the disk.
We performed a resolution test by placing a single planet on an eccentric
orbit that initially made excursions into the disk at apastron.  The evolution
of the planet's mass, eccentricity, and semi-major axis showed good agreement
between resolutions of $1.5 \times 10^5$ and $3.0 \times 10^5$ particles.  Figure \ref{density_example} shows an example of one of our runs, showing the interaction of the outer planet with the disk approximately 30 years after the scattering.

\section{RESULTS}
Following the scattering event, we evolve the system hydrodynamically 
up to the time when the separation of the planets is great enough that -- 
{\em in the absence of gas} -- they would be Hill stable. This requires a 
relatively short integration of the order of $10^4$~yr in duration. It would 
be desirable to continue the hydrodynamic simulation until all of the gas 
has been dispersed, but this is both computationally more difficult and would 
require the addition of additional physics (such as photoevaporation) describing the 
final clean-up of the gas disk. In the paper, we instead study the longer term 
evolution using N-body calculations that include an approximate treatment 
of the effects of gas disk damping on planetary eccentricity.  

\subsection{Short-term evolution}

\begin{figure}
\centering
  \plotone{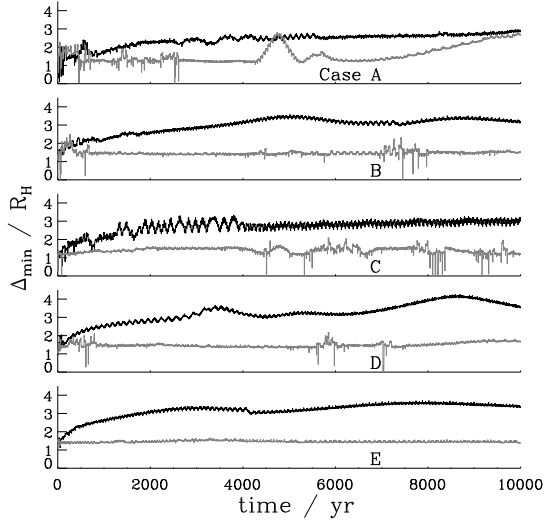}
  \caption{The minimum separation (equation \ref{delta_definition}) for each
  case in units of mutual Hill radii.  {\it Thick}: the simulations with the
  gas disk.  {\it Thin}: the same initial conditions run with no disk.}
  \label{separations}
\end{figure}

Figure~\ref{separations} shows how the separation of the planets evolves with
time in runs with and without gas. As a measure of the stability of the
systems, we calculate the minimum separation, defined by the instantaneous
values of the semi-major axes $a$ and eccentricities $e$ of the inner and
outer planets by
\begin{equation}
  \Delta_{min} = a_{out}(1-e_{out}) - a_{in}(1+e_{in}).
  \label{delta_definition}
\end{equation}
Figure~\ref{separations} shows this quantity as a fraction of the mutual Hill radius 
\begin{equation}
 R_H = \left(\frac{m_{in} + m_{out}}{3M_{\star}}\right)^{1/3} \bar{a}, 
\end{equation} 
 where $\bar{a}$ is the average of the planet's semi-major axes and
 $M_{\star}$ is the stellar mass.  For nearly-circular planets, the system is
 Hill stable when the orbital parameters satisfy $\Delta_{min} \gtrsim 3.5R_H$
 \citep{gladman93}.  Though the orbits in these simulations are moderately
 eccentric, we use this measure as a intuitive guide to short-term stability.
 \citet{marchal82} give a general condition for Hill stability, parameterized
 by a combination of the bodies' masses and the system's angular momentum and
 energy, $\zeta \propto L^2 E$.  A critical value $\zeta_{crit}$ can be found,
 above which the system is stable.  In the non-dissipative case $L$ and $E$
 are integrals of motion, and all of our simulations begin below the critical
 value.  With dissipation in the form of the gas disk, the energy and angular
 momentum of the three bodies are no longer constant, and $\zeta$ can evolve.
 In table \ref{sim_params} we give the value of $\zeta$ as a fraction of the
 critical value at the end of the simulations.  All of the dissipative cases
 have achieved Hill stability at $10^4$ yr.
  
When no gas is present, the systems do not move toward stability over the
timescale simulated, and all but one of them (case E, which is likely unstable
on a somewhat longer timescale because of its larger close approach distance)
continue to undergo strong scattering interactions.  Case A with no disk
appears to be moving toward a more stable configuration as measured by the
minimum separation, but the measure of Hill stability $\zeta$ remains as it
must at its initial value, in the unstable regime.  The simulations including
the remnant disk all follow the same general trend toward an increasing level
of stability as measured by $\Delta_{min}/R_H$.  After the first 100 years of
evolution, none of the systems with gas present undergo encounters with
$r_{enc} < R_H$, whereas the non-dissipative runs continue to undergo strong
interactions.

\begin{figure}
\centering
  \plotone{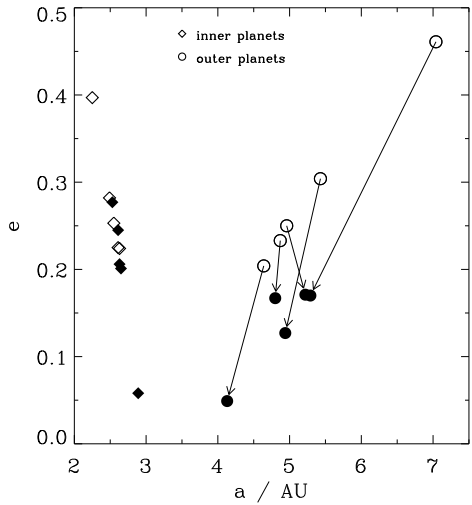}
  \caption{The mean values of the eccentricities and semi-major axes for all the simulations.  {\it Filled symbols}: simulations with disks.  {\it Open symbols}: simulations with no disks.  Arrows connect the same initial conditions with and without disks for the outer planets.}
  \label{aediff}
\end{figure}

In Figure~\ref{aediff} we plot the mean values of $a$ and $e$ over the span of the simulations for the different cases, with and without the gas disk.  While the inner planets show no clear separation in distribution, there is a consistent difference in eccentricity for the outer planets.  The arrows connect the properties of the outer planet for each case, with and without remnant disks.  In each instance, the simulation with a disk ends up at a lower eccentricity.  The effect of the disk on the inclination evolution of the systems is much less clear, with no consistent trend appearing in our simulations.

\begin{figure}
\centering
  \plotone{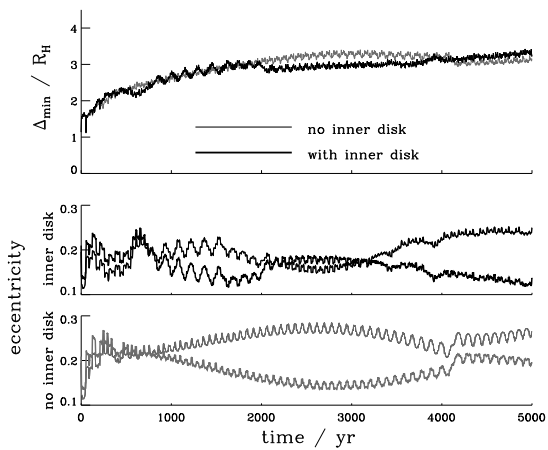}
  \caption{Comparison between case E with and without an inner disk.  {\it Top}: the minimum separation (equation \ref{delta_definition}) for the two cases.  Both have reached Hill stability by 5000 yr.  {\it Bottom two panels}: the eccentricities of both planets for each case.  The two planets are identified by thick and thin lines.  The details of the eccentricity evolution differ when an inner disk is added, though the final result in the sense of stability is very similar.}
  \label{inner_disk}
\end{figure}

We ran case E with the addition of an inner disk as well.  Case E was chosen because its inner
planet had the smallest periastron post-scattering, so that any effect of accretion from the inner
disk would be maximized.  The total angular momentum of the inner disk we simulated is $\sim$4\%
of the inner planet's orbital angular momentum, and we would expect the change in eccentricity
associated with accreting this material to be small.  In figure \ref{inner_disk} we compare the
evolution of the orbits with the different disk profiles.  The details of the semi-major axis and
eccentricity evolution between the two realizations are different, as the initial evolution of the
inner planet is changed and sets the system on an altered path.  The separation measure
$\Delta_{min}$ shows a very similar evolution, however, with the same trend toward stability.  As
in the case with no inner disk, the outer planet accretes significantly from the exterior disk and
circularizes.  At 5000 years after the close encounter, the system has reached Hill stability
($\zeta / \zeta_{crit} = 1.033$) and the inner disk has been completely disrupted.  The further
evolution of the system will be driven by accretion onto the outer planet.  We conclude that the
presence of a depleted inner disk is unlikely to significantly affect the results of our
simulations.

In contrast to dynamical friction, resonant interaction with the gas disk, or ejection of
planetesimals, the stabilizing influence of these gas disks is due to the accretion of material by
the outer planet.  As it makes excursions into the disk near its apastron, it is accreting
material with higher specific angular momentum, which causes its eccentricity to decrease.  This
effect is therefore dependent on enough material being present in the disk.  Tests with 0.1 $M_J$
of disk material showed very little difference from the runs with no gas present.  By the end of
the $10^4$ yr run of the simulations the mass remaining in the disk has been reduced to $\sim 0.2
- 0.4M_J$, while the outer planet has gained mass ($m_{out}\sim 1.5M_J$, see Table
\ref{sim_params}). Due to the expense of the hydrodynamic simulations, we are unable to extend
the runs over disk-driven migration timescales. We cannot directly guarantee that further viscous 
evolution of the disk will not push the planets back toward instability, and restart the 
whole cycle. However, this possibility is unlikely. Once the disk mass drops significantly 
below the planet mass the rate of migration is suppressed, though the exact degree 
depends upon the disk structure (e.g. Armitage 2007). The damage done to the outer 
disk by the initial scattering and subsequent accretion will have reduced the migration 
rate by a factor of several, making it unlikely 
that the disk will be able to push the system back toward instability during its remaining
lifetime. A second round of instability is even less likely given the fact that 
photoevaporation of a sub-Jupiter mass gas disk by direct stellar illumination 
occurs on a rapid time scale \citep{alexander06}.

\begin{figure}
\centering
  \plotone{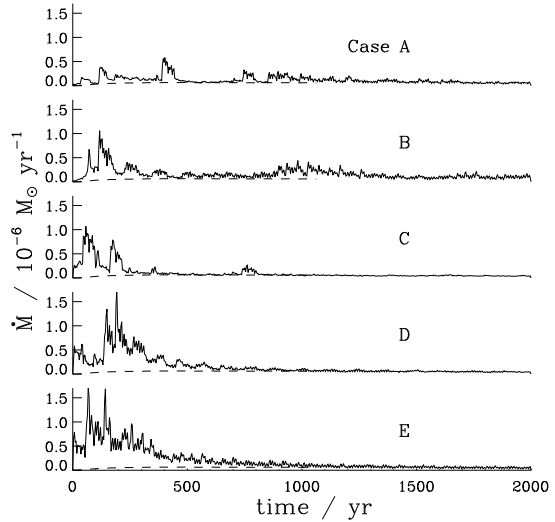}
  \caption{The accretion rate of the outer planet for each case, for the first
  2000 years of the simulation.  Also plotted ({\it dashed}) is a reference accretion rate in
  the absence of scattering.}
  \label{acc_rates}
\end{figure}

The accretion levels during the outer planet's initial interactions with the
disk, when its eccentricity is damped most rapidly, are quite high.  In figure
\ref{acc_rates} we show these rates for the first 2000 years of the
simulations, i.e. immediately after the first scattering event.  The cases
with the weakest initial encounters exhibited deeper penetration into the
disk, and have higher accretion rates than the cases with stronger
interactions.  With the exception of cases A and B, which show secondary
bursts of accretion at $\sim 1000$ years, the main accretion onto the planets
occurs over the first 500 years of the simulations.  As a reference accretion
rate, we ran a simulation with a single planet orbiting at the 2:1 resonance
with the inner edge of the disk for 1000 yr.  This rate is plotted in figure
\ref{acc_rates} as well, and has a mean value $\sim 5\times 10^{-8}$ \msuns
yr$^{-1}$.

\subsection{Long-term evolution}
We terminated our SPH simulations with the inner planet on an eccentric orbit
(typically $e \sim 0.2-0.3$) and the outer planet still interacting with the
remnant gaseous disk, though much more weakly than immediately after the scattering encounter.  The lifetime of the remnant gaseous disk is uncertain:
$10^{5-6}$ yr is a reasonable estimate based on models of dissipation via photoevaporation
\citep{clarke01,alexander06} and observational constraints \citep{simon95,wolk96}.  Given that the outer planet is experiencing
orbital damping and is also gravitationally interacting with the inner planet,
an interesting question is whether the inner planet's eccentricity can be
decreased.  We therefore investigate the long-term dynamical evolution of
these systems.

We model the system using the hybrid integrator {\tt Mercury} \citep{chambers99}. 
The physical scenario that we have in mind is that the eccentricity of the 
outer planet will be damped on long time scales via resonant interaction 
with the remaining gas disk, once it has settled down following the 
perturbation induced by the scattering. The exact form of the damping is 
not likely to be critical, so for simplicity we use a method that we have tested 
for other applications and apply 
a drag force to the outer planet in the form of a headwind which is
proportional to the difference in velocity between the planet and the gas
\citep[which is assumed to be partially pressure supported, as in][]{thommes03}. 
This force is therefore equivalent to aerodynamic drag \citep{adachi76}.  
For our purposes we treat the strength of the damping as a free parameter, and calibrate  
the resulting eccentricity damping to match the damping of the outer planet in our SPH
simulations.  The effect of this force is to damp $e$ on a timescale of a few
times $10^4$ years with little damping of $a$, as long as $e$ is less than
$\sim 0.2$.  We therefore believe that our implementation is a reasonable
approximation.

\begin{figure}
\centering
  \plotone{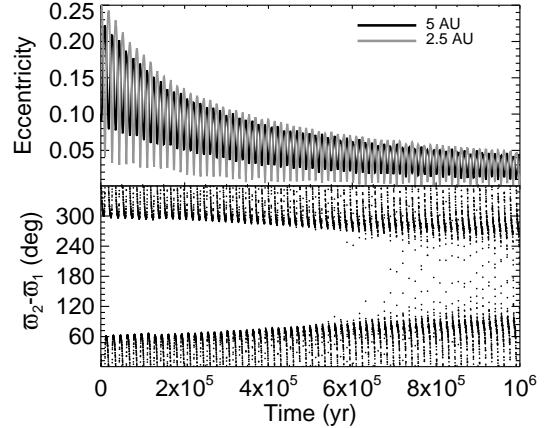}
  \caption{Long-term orbital evolution of two Jupiter-mass planets with a drag force
being applied to the outer planet.  {\it Top panel}: Eccentricities of the inner
({\it grey line}) and outer ({\it black line}) planets in time.  {\it Bottom panel}: the
relative apsidal angle, i.e. the angle between the longitudes of pericenter of
the two planets, $\varpi_1-\varpi_2$.}
  \label{longterm_const}
\end{figure}

Figure \ref{longterm_const} shows the evolution of two Jupiter-mass planets for a case that
started with the inner planet with $a$ = 2.5 AU and $e = 0.2$ and the outer
planet at 5 AU with $e = 0.1$.  The outer planet's eccentricity is quickly
excited by the inner planet and the two planets undergo regular secular
oscillations with a period of about 15,000 years.  The initial amplitude of
eccentricity oscillations is close to the inner planet's starting
eccentricity, and the two values remain out of phase.  The damping decreases
the eccentricities of both planets in time, but does not affect the secular
frequency.  At the end of the simulation, the inner/outer planets have median
eccentricities of 0.036/0.032.  The damping also affected the planets' apsidal
aligment, as measured by the difference in the longitudes of pericenter, i.e.,
$\varpi_1 - \varpi_2$.  The two planets quickly became locked in libration
about apsidal alignment ($\varpi_1 - \varpi_2 = 0$), but the libration
amplitude increased in time until the nodes began to occasionally circulate
after $\sim 6 \times 10^5$ years.  For all the configurations that we tested,
evolution moved the planets away from apsidal alignment and toward
anti-alignment.  For cases in which the outer planet started with $e=0$, the
system quickly tended toward anti-alignment and the amplitude of libration
decreased steadily to from $>90 \deg$ at early times to less than $50 \deg$
after 1 Myr \citep[see][]{chiang02}.

\begin{figure}
\centering
  \plotone{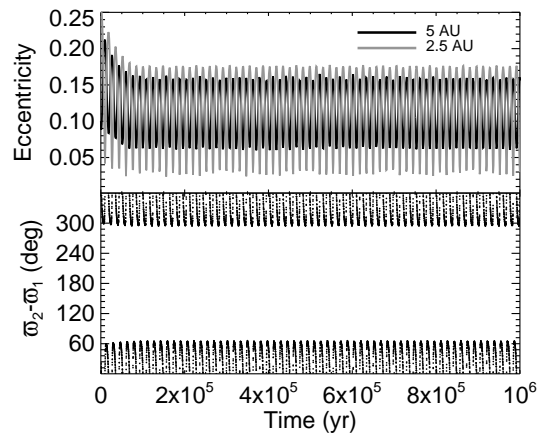}
  \caption{Long-term orbital evolution of two Jupiter-mass planets with a time-dependent
drag force applied to the outer planet.  The drag force decreases linearly to
zero after $10^5$ years and no drag is applied for the rest of the
simulation.  }
  \label{longterm_var}
\end{figure}

The finite lifetime of the gaseous disk may limit the damping of the
eccentricity of the outer, and therefore also inner, planet.  The case from
figure \ref{longterm_const} assumed constant damping from the disk over 1 Myr.  However, the outer
disk is being depleted by both accretion onto the outer planet and
photoevaporation.  Figure \ref{longterm_var} shows the evolution of the same two
planets from figure \ref{longterm_const} but assuming the damping dissipated linearly from its
default value to zero after $10^5$ years, a more reasonable value for the length of
the final stage of disk dissipation \citep[e.g.][]{clarke01,alexander06}.
Eccentricities of the planets in figure \ref{longterm_var} are far less damped than in
figure \ref{longterm_const}: the median final eccentricities of the inner/outer planets are
0.127/0.12.  Note that for the case of no damping, the median final
eccentricities of the inner/outer planets are 0.24/0.14.  Thus, even a short
period of damping can reduce eccentricities.  Note also that the weaker
damping has slightly increased the libration amplitude but not enough to cause
the apses to circulate.  We regard this second setup, with a limited amount of damping, as the more realistic scenario.  The remnant disk mass is reduced by the initial planet-disk interactions, and estimates of the photoevaporation timescale are closer to $10^5$ than $10^6$ yr. 

The known extra-solar multi-planet systems are preferentially located close to
the separatrix between apsidal circulation and libration \citep{barnes06}.  Near-separatrix behavior is characterized by one of the planets
periodically reaching an $e=0$ state \citep[e.g.][]{ford05}.  The strong
damping experienced by the system in figure \ref{longterm_const} has driven the system from
libration to a near-separatrix state, whereas the system from figure \ref{longterm_var} is
steadily librating.  For cases which started with a zero eccentricity outer
planet, the situation is reversed: simulations that experienced only a small
amount of damping remained close to the separatrix but those that were
strongly damped were driven to low-amplitude apsidal libration about
anti-alignment.  Thus, interactions between planets during this phase of a
low-mass, remant gas disk may play an important role in establishing the final
apsidal alignment of planetary systems.

\section{DISCUSSION}
The expense of the hydrodynamic simulations means that we have been able to 
run only a handful of realizations of scattering in the presence of gas, and as a 
consequence it is unwise to draw definitive conclusions. Nonetheless we find, 
as expected, that remnant disks that have masses comparable to the masses of the 
giant planets {\em do} influence the outcome of gravitational scattering, primarily 
via eccentricity damping that occurs due to accretion when planets intrude upon 
the gas disk. If, as we have assumed, the remnant gas lies exterior to the orbit 
of the outermost planet, damping when the planet enters the disk tends to increase the 
planetary separation and stabilize the system against close encounters and 
possible ejection of one of the planets. The stabilization occurs rapidly enough 
that we were able to evolve the systems with gas up to the point where they 
would plausibly be Hill stable.

The observed eccentricity distribution of extrasolar planets\footnote{A detailed 
study of possible selection biases in the Keck planet search \citep{cumming08} shows that the survey is complete for velocity amplitudes $K > 20 \ {\rm ms}^{-1}$ 
and $e < 0.6$. The ``observed" distribution is therefore a reasonable proxy for the 
true distribution in the sense that neither low nor moderately high $e$ objects are 
systematically missed. There may, however, be some upward bias to the derived 
eccentricities of detected planets in cases where the measurement signal to 
noise is relatively low \citep{shen08}.} matches the theoretically 
predicted one {\em from simulations that do not include gas} \citep{chatterjee07,juric07}. 
If gas is implicated in driving planetary systems toward instability, its effect on the 
subsequent dynamics cannot be so strong as to ruin this agreement. We find that this 
is plausible. Specifically, models computed with an exterior disk whose mass equals 
that of the planets yield inner planet eccentricities that are comparable to those 
achieved in the absence of gas. Since radial velocity surveys favor the detection 
of planets at smaller orbital radii, the observed sample today is likely dominated 
by planets which would have been unaffected by gas during the scattering process. 
Conversely, we find that the eccentricity of the outer planet is significantly 
damped by accretion of gas once it enter the disk. The dynamical dominance of 
accretion over other damping mechanisms is likely the most robust result of this 
paper. 

Although we have only considered one specific scattering scenario, it seems 
plausible that the generic effect of residual gas disks will be to increase the 
fraction of retained rather than ejected planets following scattering. If so, 
analysis of the abundance of multiple planets ought to provide the strongest 
constraints on models of this type. If the specific model for 
migration and scattering we have used were actually commonplace in extrasolar 
planetary systems, we would expect to find a high fraction of multiple planet systems 
in which the outer planet had a lower eccentricity than the inner planet. 
Observationally, it is known that there are hints for additional companions 
in many systems that host known planets \citep{wright07}, but the 
quantitative statistical analysis necessary to place limits on their masses 
and periods (analogous to the work of \citealt{cumming08} for single planets) 
has not, to our knowledge, been attempted. Our impression is that while 
our toy model -- in which the hypothetical outer planet has the same mass 
as the detected inner one -- could already be ruled out observationally, 
a more realistic scenario in which the outer planet was modestly less 
massive probably remains viable. More extensive theoretical work will be 
needed to quantify the extent to which residual gas disks increase the predicted fraction 
of multiple systems after scattering.

On Myr timescales, the inner planet's eccentricity can be damped indirectly by
the remnant gaseous disk via excitation of the outer planet and subsequent
dissipation.  The lifetime of the remnant disk is uncertain, but is probably
$\sim 10^5$ years \citep{simon95,wolk96,clarke01,alexander06}.  Given the large
eccentricity of the inner planet and the relative proximity of the outer
planet, there is strong secular coupling and eccentricity oscillations are
induced.  The damping of the outer planet by the remnant gas disk tends to
push the system away from apsidal alignment and toward anti-alignment; the
amount of damping depends on the disk lifetime.  Depending on the starting
configuration, strong damping can produce systems that are near the apsidal
separatrix \citep{barnes06}, circulating, or librating about
anti-alignment.  However, strong damping should not produce systems that
librate about alignment.  In cases with relatively little damping, the apsidal
alignment is weakly affected, although eccentricities can be modestly reduced
(see figure \ref{longterm_var}).

Finally, we note that the accretion rate onto the planet when the planet first
enters the disk is rather high. This gas will, in reality, join a
circumplanetary disk, and how quickly it is accreted by the planet will depend
upon the evolution of that disk.  However, it is clear that the luminosity of
the planet could be large due to the implied late burst of accretion. Naively
estimating the accretion luminosity as $L_{acc} \sim G M_p \dot{M} / R_p$, we
find that for Jovian parameters the luminosity over the first 500~yr is of the
order of $10^{32}$-$10^{33}$ erg s$^{-1}$. Since the accreted gas has angular 
momentum, the mass inflow onto the planet would be buffered by  
a circumplanetary disk, though only modestly since the viscous timescale of 
such disks is typically of the order of $10^3 \ {\rm yr}$ or less \citep{canup02}. Unlike the main envelope accretion
event that occurs in core accretion models for giant planet formation, this
late burst of accretion is in principle observable (as a near-IR excess without 
matching stellar accretion signatures) since the accretion
luminosity of the circumstellar disk at this epoch is small. If we assume that
late accretion is detectable for $10^3 \ {\rm yr}$, and occurs in 10-100\% of
young stars, then we would need to survey $10^4$ to $10^5$ systems to have a
chance of directly observing the last stage of giant planet formation.
Surveys of this magnitude are well beyond current capabilities, but are not
inconceivable in the future. 

\acknowledgements

We thank the referee for a helpful report, and in particular for noting the 
possible importance of secular damping on longer timescales. 
This research was supported by NASA under grants NNA04CC11A to the CU Center
for Astrobiology, NNG04GL01G and NNX07AH08G from the Astrophysics Theory
Programs, and by the NSF under grant AST~0407040.  S.N.R. was supported by an
appointment to the NASA Postdoctoral Program at the University of Colorado
Astrobiology Center, administered by Oak Ridge Associated Universities through
a contract with NASA. P.J.A. acknowledges the hospitality of the Physics and 
Astronomy Department at UCLA, where this work was completed.

%\bibliography{scatteringrefs}

\end{document}